\documentclass[twoside]{llncs}
\usepackage[T1]{fontenc}
\usepackage{enumerate}
\usepackage[shortlabels]{enumitem}
\usepackage{xcolor}
\usepackage{graphicx}
\usepackage{booktabs}
\usepackage[normalem]{ulem}
\useunder{\uline}{\ul}{}

\begin{document}


%
\title{Accountability of Robust and Reliable AI-Enabled Systems: A Preliminary Study and Roadmap}
%
%
\author{Filippo Scaramuzza\\supervised by\\Damian A. Tamburri and Willem-Jan Van Den Heuvel}
\authorrunning{F. Scaramuzza et al.}
%
\institute{Jheronimus Academy of Data Science, Tilburg University, 's-Hertogenbosch, NL \\
\email{f.scaramuzza@tilburguniversity.edu\\d.a.tamburri@tue.nl\\w.j.a.m.v.d.heuvel@jads.nl}}
\maketitle              

\begin{abstract}

This vision paper presents initial research on assessing the robustness and reliability of AI-enabled systems, and key factors in ensuring their safety and effectiveness in practical applications, including a focus on accountability. By exploring evolving definitions of these concepts and reviewing current literature, the study highlights major challenges and approaches in the field. A case study is used to illustrate real-world applications, emphasizing the need for innovative testing solutions. The incorporation of accountability is crucial for building trust and ensuring responsible AI development. The paper outlines potential future research directions and identifies existing gaps, positioning robustness, reliability, and accountability as vital areas for the development of trustworthy AI systems of the future.

\end{abstract}
\section{Introduction and Problem Statement}

AI-enabled enterprise applications are transforming industries with new levels of automation and intelligence. Trust in these systems is now essential, spanning physical, emotional, digital, financial, and ethical domains. The rapid development of AI offers significant economic and social benefits across various sectors, including transportation, finance, medicine, security, and entertainment \cite{liTrustworthyAIPrinciples2023a}. However, breaches of stakeholders' trust can lead to severe societal consequences, such as biased hiring decisions \cite{HelpWantedExamination} and safety risks \cite{boudetteItHappenedFast2021}. 
This increasing reliance on AI ties into AI governance, which encompasses the rules, ethical standards, and frameworks guiding AI development and deployment \cite{AIGovernance}. Businesses are beginning to realize numerous benefits from AI technologies, such as improved customer relationships and enhanced operational efficiency \cite{hoiResponsibleAITrusted2023}. According to MarketsandMarkets, the AI market, valued at \$86.9 billion in 2022, is expected to grow to \$407 billion by 2027, with a compound annual growth rate (CAGR) of 36.2\% \cite{hoiResponsibleAITrusted2023}.


Historically, the terms \textit{robustness} and \textit{reliability} have had distinct meanings. Their definitions have evolved over time, with three notable interpretations ordered by age. The \textit{Shorter Oxford English Dictionary} (SOED) \cite{pressNewShorterOxford1997}, \textit{Steve McConnell} (in Code Complete, CC) \cite{mcconnellCodeComplete2004}, and \textit{Bertrand Meyer} (in Object-Oriented Software Construction, OOSC) \cite{ObjectOrientedSoftwareConstruction} provide the definitions shown in Table \ref{table:rr_definitions}.

\vspace{-4mm}

\begin{table}[]
\resizebox{\textwidth}{!}{%
\begin{tabular}{@{}lp{5cm}p{5cm}p{5cm}@{}}
\toprule
                     & \textbf{SOED}                                                                                                                                                                        & \textbf{CC}                                                                                                                                      & \textbf{OOSC}                                                                  \\ \midrule
\textbf{Robustness}  & strong and sturdy in physique or construction. Involving or requiring bodily or mental strength or hardiness. Strong, vigorous, healthy, not readily damaged or weakened. & The degree to which a system continues to function in the presence of invalid inputs or stressful environmental conditions.                      & The ability of software systems to react appropriately to abnormal conditions. \\
\textbf{Reliability} & That which may be relied on, in which reliance or confidence may be put, trustworthy, safe, sure.                                                                                   & The ability of a system to perform its requested functions under stated conditions whenever required - having a long mean time between failures. & A concern encompassing correctness and robustness.                             \\ \bottomrule
\end{tabular}%
}
\caption{Formal definitions of robustness and reliability ordered by age.}
\label{table:rr_definitions}
\end{table}

\vspace{-8mm}


As AI systems are increasingly integrated into critical applications, accountability becomes crucial \cite{krollOutliningTraceabilityPrinciple2021,nushiAccountableAIHybrid2018}. It involves clear responsibility for system behavior, auditability, and mechanisms for recourse in case of failures \cite{hutchinsonAccountabilityMachineLearning2021}. Reliability in AI includes attributes like generalizability and adaptability, while robustness addresses resilience under adversarial conditions \cite{jiReliabilityRobustnessResilience2023,ModelRobustnessBuilding}. Distinguishing these concepts is vital for developing AI systems that ensure performance, safety and accountability in practical applications.

\section{Literature Review and Case Study}
This section delves into the literature that offers a foundation for understanding these concepts, with an emphasis on practical approaches and emerging tools. Additionally, an industry case study is presented to demonstrate real-world challenges in developing accountable, reliable and robust AI systems, providing insights into testing, implementation, and performance evaluation.

\subsection{Literature Review}

Hong et al. \cite{hongStatisticalPerspectivesReliability2023} introduce ``SMART’’ (Structure, Metrics, Analysis, Reliability Assessment, and Test Planning) for evaluating AI reliability over time. Blood et al. \cite{bloodReliabilityAssuranceAI2023} integrate traditional tools like FMEA and HALT with modern approaches like STPA for AI systems.
Kroll \cite{krollOutliningTraceabilityPrinciple2021} stresses that accountability requires understanding a system’s function, creation, and purpose. This means adopting frameworks that support traceability, documenting design decisions, data sources, and algorithmic choices.
Schmelczer et al. \cite{schmelczerTrustworthyRobustAI2023a} propose ``GreatAI’’ for transitioning AI from prototype to production, while Chen et al. \cite{chenAIMaintenanceRobustness2023} emphasize lifecycle oversight. Wozniak et al. \cite{wozniakRobustnessTestingIndustrial2023} highlight data augmentation for robustness, but Hutchinson et al. \cite{hutchinsonAccountabilityMachineLearning2021} note the accountability implications. 
Raji et al. \cite{rajiClosingAIAccountability2020b} introduce a framework for internal algorithmic auditing to ensure that AI systems are developed in alignment with ethical expectations and standards.
In MLOps \cite{kreuzbergerMachineLearningOperations2023a}, Laato et al. \cite{laatoAIGovernanceSystem2022} introduce AI governance models, while Ashmore et al. \cite{ashmoreAssuringMachineLearning2021} and Li et al. \cite{liTrustworthyAIPrinciples2023a} embed Trustworthy AI principles. Matsui et al. \cite{matsuiMLOpsGuideIts2023} and Billeter et al. \cite{billeterMLOpsEnablerTrustworthy2024} provide practical MLOps guidelines.
Accountability in MLOps involves clear agreements on ML service quality and performance. Truong and Nguyen \cite{truongQoA4MLFrameworkSupporting2021} discuss ML contracts beyond traditional SLAs, and Ahmed et al. \cite{ahmedDesignContractDeep2023} propose frameworks for specifying and enforcing preconditions and postconditions to prevent bugs.

These studies highlight the importance of Trustworthy AI, a key focus of this PhD project, supported by Deloitte Netherlands\footnote{\url{https://www.deloitte.com/nl/nl.html}}.

\subsection{Industry Case Study} A European tech lab developed a chatbot system using Large Language Models (LLMs) and Retrieval-Augmented Generation (RAG) to balance user freedom with author control while adapting to unexpected inputs in educational and entertainment contexts. The project faced testing challenges, particularly with inconsistent LLM results. Accountability and responsibility quickly emerged as essential considerations, particularly due to the system's potential impact on end users. Developers addressed this through repeated unit tests and extensive benchmarks, which required significant expertise and cost \cite{231214231BuildingYoura}. A key question remained: when is performance ``good enough''? Engineers and developers from the project participated in a survey, and their responses are summarized below:

\begin{enumerate} 
    \item \textit{What strategies would you have implemented differently to improve robustness and reliability, knowing the outcomes?} --- Developers recommended utilizing pre-built frameworks for API interactions and integrating RAG to improve data retrieval. They emphasized the need for fine-tuning models to handle out-of-context inputs and acknowledged the challenges in accurately measuring model reliability, especially in terms of accountability to end users.
    \item \textit{Describe a specific instance where the robustness of an AI system was tested in production. What lessons did you learn?} --- In a benchmarking scenario, the system showed robustness on simple tasks but struggled with more complex datasets. The most valuable lesson learned was the importance of prompt engineering coupled with recognizing that robustness alone does not guarantee responsible and accountable deployment.
    \item \textit{Which research papers or studies have shaped your understanding of AI system reliability?} --- Developers pointed to areas like RAG, information grounding, and prompt engineering, though they did not cite specific papers.
    \item \textit{What emerging research topics will be critical for advancing AI reliability?} --- Key areas included improving RAG to reduce hallucinations, enhancing prompt engineering, and developing explainable AI models. The need for better data quality, increased explainability, and well-defined accountability and responsibility measures was emphasized.
\end{enumerate}

\subsection{Identified Limits in the State of the Art}
There is a lack of comprehensive studies on the reliability and robustness of AI systems and a gap between industry and academia in this area. Few surveys focus on AI robustness, and there's no broad categorization of AI models in industrial applications. Trustworthiness in AI systems also lacks a formal definition beyond neural networks and LLMs. This absence of clear definitions and evaluations can hinder the development of accountable AI systems.
Empirical evaluations of Trustworthy AI principles in current systems are limited, underlining the need for further research in this field and MLOps.
Survey results highlight the value of pre-built frameworks and libraries for AI development, focusing on RAG and multi-agent systems. Robustness testing stressed the importance of clear prompts, error correction, and user input filtering. Emerging areas like prompt engineering, real-time data retrieval, and explainable AI are seen as key to improving AI reliability. However, the absence of specific academic citations reveals a gap in referencing relevant research.

\section{Envisioned Research Plan}
As a starting point for the research, here are outlined a few research questions that can be used to further explore these topics:
\begin{enumerate}[start=0, label={\bfseries RQ\arabic*)}, leftmargin=1.5cm]
    \item What are the advances and challenges in creating \textbf{data pipelines} that support the \textbf{development of accountable and trustworthy AI}, and how do these pipelines influence the robustness of complex AI models, like multi-agent systems?
    \item What are the key components of a \textbf{framework} designed to \textbf{ensure safe and accountable AI}, and how can these components be integrated to \textbf{create reliable AI systems}?
    \item What mechanisms can be implemented to enhance the \textbf{verifiability of claims} regarding the safety, security, and fairness of AI systems, and how do these mechanisms contribute to the overall accountability of AI development and deployment?
    \item  How can \textbf{MLOps practices} be effectively integrated with \textbf{Trustworthy Artificial Intelligence principles} to optimize the lifecycle management of AI systems, ensuring their ethical \textbf{deployment and operational integrity} and how can accountability be operationalized within this integrated framework?
\end{enumerate}

\section{Conclusions}
This study highlights the importance of robustness, reliability, and accountability in AI-enabled systems, where accountability, in particular, is essential for maintaining trust in deployment. Embedding comprehensive auditing practices throughout the AI lifecycle is key to guarding against adversarial threats and upholding data quality. Techniques like adversarial training, sensitivity reduction, and expert-reviewed data are foundational for achieving reliable performance across diverse applications, while clear accountability structures ensure responsible outcomes.

The industry case study further demonstrates gaps between current practices and the complex demands of real-world AI, pointing to a pressing need for frameworks that harmonize accountability with operational needs. Addressing these gaps is crucial for advancing trustworthy, accountable AI systems across various sectors.

\vspace{-1mm}


%
%
%
\bibliographystyle{unsrt}
\bibliography{bibliography}
\end{document}